\documentclass[12pt]{article}

\usepackage{graphicx}% Include figure files

\def \lsim{\mathrel{\vcenter
     {\hbox{$<$}\nointerlineskip\hbox{$\sim$}}}}

\newcommand{\beq}{\begin{equation}}
\newcommand{\eeq}{\end{equation}}
\newcommand{\beqa}{\begin{eqnarray}}
\newcommand{\eeqa}{\end{eqnarray}}
\newcommand{\beqar}{\begin{eqnarray*}}
\newcommand{\eeqar}{\end{eqnarray*}}

\begin{document}
\thispagestyle{empty}

\hfill{\sc UG-FT-236/08}

\vspace*{-2mm}
\hfill{\sc CAFPE-106/08}

\vspace{32pt}
\begin{center}
\centerline{\textbf{\Large
A minimal Little Higgs model}}

\vspace{40pt}

Roberto Barcel\'o and Manuel Masip
\vspace{12pt}

\textit{
CAFPE and Departamento de F{\'\i}sica Te\'orica y del
Cosmos}\\ \textit{Universidad de Granada, E-18071, Granada, Spain}\\
\vspace{16pt}
\texttt{barcelo@correo.ugr.es, masip@ugr.es}
\end{center}

\vspace{40pt}

\date{\today}% It is always \today, today,

\begin{abstract}

We discuss a Little Higgs scenario that introduces 
below the TeV scale just the
two minimal ingredients of these models, a vectorlike
$T$ quark and a singlet component (implying anomalous 
couplings) in the Higgs field, together with a pseudoscalar
singlet $\eta$.
In the model, which is a variation of Schmaltz's 
{\it simplest} Little Higgs model, all the extra 
vector bosons are much heavier than the $T$ quark. 
In the Yukawa sector the global symmetry is {\it approximate},
implying a single large coupling per flavour, 
whereas in the scalar sector it is only broken at 
the loop level. 
We obtain the one-loop effective potential
and show that it provides acceptable masses for the 
Higgs $h$ and
for the singlet $\eta$ with no need for 
an extra $\mu$ term. We find that $m_\eta$ can be
larger than $m_h/2$, which would 
forbid the (otherwise dominant) 
decay mode $h\rightarrow \eta\eta$.

%\pacs{12.60.Fr}% PACS, the Physics and Astronomy
%\keywords{Neutrino Masses, Higgs Physics, Beyond Standard Model}
                              %Use showkeys class option if keyword
\end{abstract}

\newpage

\section{Introduction}

The Large Hadron Collider (LHC) at CERN should reveal in the
next few years the value and the {\it nature} of the 
Higgs boson mass. The confirmation of the standard model (SM) 
Higgs sector, with no signs of new physics, 
would certainly be a very interesting 
possibility \cite{Agrawal:1997gf,Arkani-Hamed:2004fb}. Most of the 
community, however, expects (or hopes)
that there is a dynamical explanation to the hierarchy problem, and that
this explanation will become apparent at the LHC. 

One of the possibilities that recently has attracted most 
attention is the Little Higgs (LH) scenario 
\cite{Georgi:1974yw,Arkani-Hamed:2001nc,Arkani-Hamed:2002qy}. 
Its main motivation has been 
{\it experimental}: the absence of any deviations to the SM predictions 
in all precision data. Supersymmetry (SUSY), 
technicolor or the 
presence of extra dimensions could rise the natural 
cutoff of the SM and define a theory that is  
consistent up to the fundamental scale. To do the work,
however, this new physics should appear below the TeV, whereas
the experiments
seem to bound it to be above 5--10 TeV \cite{Erler:2004nh}.
LH models would release this tension by providing an explanation
for the gap between the electroweak (EW) scale and the 
scale of the new physics. It is not that LH
does not imply physics beyond the SM (it does), but being its 
objective and its structure more simple 
it tends to be more consistent with the data than these other 
fundamental mechanisms. The LH idea of the Higgs as a 
pseudo-Goldstone boson (pseudo-GB) of a broken symmetry could be 
{\it incorporated} into a SUSY 
\cite{Roy:2005hg,Berezhiani:2005pb,Csaki:2005fc,Bai:2007tv} 
or a strongly
interacting theory \cite{Contino:2003ve,Giudice:2007fh}
to explain the {\it little} hierarchy between
the Higgs vacuum expectation value (VEV) and the SUSY breaking
scale or the mass of the composite states. 
The most important consequence is then that it would 
describe all the new physics to be explored at the LHC.

In this paper we consider a variation of Schmaltz's model 
\cite{Schmaltz:2004de}
based on a $SU(3)\times SU(3)$ global symmetry, the so called 
{\it simplest} LH model (see 
\cite{Perelstein:2005ka,Schmaltz:2005ky} for a review). 
This model includes two VEVs, $f_{1,2}$, that break
the global symmetry giving mass to a vectorlike $T$ quark 
and to several nonstandard gauge bosons. The modifications
that we propose are the following. First, we separate by a 
sizeable factor the two VEVs, $f_1\approx 0.1 f_2$.
This region of the parameter space, identified 
by other authors 
\cite{Casas:2005ev,Han:2005dz,Marandella:2005wd} 
as the most promising
from a phenomenological point of view, implies a
$T$ quark that can be light (and cancels quadratic
corrections) while the extra vector bosons are heavier (and
consistent with precision EW data). Second,  we also 
change the usual structure of the couplings 
in the top-quark sector. Instead of two similar Yukawa 
couplings that break {\it collectively} the global
symmetry, we propose that the symmetry is approximate,
{\it i.e.}, there is one unsuppressed (symmetric) coupling 
and the rest of them break the symmetry but are smaller
by a factor of (at least) $\approx 0.1$. This has important 
implications in the one-loop Higgs potential. In particular,
the model is consistent without the need of an extra $\mu$-term

The model includes a lighter scale $f_1$ that can be close to
the electroweak (EW) scale, $v/\sqrt{2}$. As a consequence, the 
non-linear expansion of the Higgs field 
requieres a treatment beyond the usual one at first order in
$v/f_1$. Here we sum the whole series and express the result
as a function of the sine of the ratio $v/(\sqrt{2}f_1)$. 
The model is minimal in the sense that, 
in addition to the $T$ quark, only 
a neutral scalar field $r_1$ gets its mass at $f_1$: the rest of
scalars, the extra gauge bosons and the nonstandard fermions 
(up-type quarks and right handed neutrinos) 
get masses of the order of the larger scale $f_2$.
Below $f_1$ one is left in the scalar sector with the
pseudo-GBs of the global symmetry: the
SM Higgs fields plus the extra CP-odd singlet 
$\eta$ \cite{Kilian:2004pp,Kilian:2006eh}. Therefore, 
all the exotic physics that the LHC would
face below 3 TeV would be a Higgs boson with anomalous 
gauge and Yukawa couplings 
(see next section), the neutral scalars $\eta$ and 
(possibly) $r_1$, and the extra $T$ quark. 
We find the one-loop effective potential and show that this
setting naturally provides an acceptable EW
symmetry breaking with large masses for the Higgs 
and $\eta$. We also find that 
the singlet $\eta$ may be heavy enough to close
the interesting decay channel $h\rightarrow \eta \eta$ 
\cite{Cheung:2006nk,Cheung:2007sva}.

\section{Little Higgs or extra singlet model?}

Let us start reviewing the model in some 
detail \cite{Schmaltz:2004de}. The scalar sector 
contains two triplets, $\phi_1$ and $\phi_2$, of a global
$SU(3)_1\times SU(3)_2$ symmetry:
\beq
\phi_1\rightarrow e^{i\theta^a_1 T^a} \phi_1\;,\;\;\;\;
\phi_2\rightarrow e^{i\theta^a_2 T^a} \phi_2\;,
\eeq
where $T^a$ are the generators of $SU(3)$. 
It is then assumed that these triplets get 
VEVs $f_{1,2}$ and break the global symmetry
to $SU(2)_1\times SU(2)_2$. The spectrum of scalar fields
at this scale will consist of 10 massless modes (the GBs of the 
broken symmetry) plus two massive fields (with masses
of order $f_1$ and $f_2$). If one combination of the two 
global $SU(3)$ is made local, some of the GBs
will be {\it eaten} by massive gauge bosons and the rest
will define the EW scalar sector. 

In particular, if the two VEVs are 
\beq
\langle \phi_{1}\rangle = 
\left(\begin{array}{c} 0 \\ 0 \\ f_1  \end{array}\right)\; , \;\;\;\;
\langle \phi_{2}\rangle = 
\left(\begin{array}{c} 0 \\ 0 \\ f_2  \end{array}\right)\;,
\eeq
and the diagonal combination of $SU(3)_1\times SU(3)_2$ is local,
\beq
\phi_{1(2)}\rightarrow e^{i\theta^a(x) T^a} \phi_{1(2)}\;,
\eeq
then the VEVs will break the local $SU(3)\times U(1)_\chi$ to 
the standard
$SU(2)_L\times U(1)_Y$, a process that takes 5 GBs.
The other 5 GBs (the complex doublet $(h^0\; h^-)$ and a CP-odd 
singlet $\eta$) can be parametrized 
non-linearly \cite{Coleman:1969sm}:
\beq
\phi_{1}= e^{+ i\; {f_2\over f_1} \Theta }
\left(\begin{array}{c} 0 \\  0 \\ f_1+
\displaystyle{r_1\over \sqrt{2}}
\end{array}\right)\;,\;\;\;\;
\phi_{2}= e^{- i\;{f_1\over f_2} \Theta}
\left(\begin{array}{c} 0 \\  0 \\ f_2+
\displaystyle{r_2\over \sqrt{2}}
\end{array}\right)\;,
\label{paramet}
\eeq
where 
\beq
\Theta= {1\over f}\;
\left(\begin{array}{ccc} \eta/\sqrt{2} & 0 &h^0 \\
0& \eta/\sqrt{2} & h^- \\
h^{0\dagger} & h^+ & \eta /\sqrt{2} \end{array}\right)\;,
\label{theta}
\eeq
$f=\sqrt{f_1^2+f_2^2}$, and $r_{1(2)}$ is a scalar
with mass of order $f_{1(2)}$.

If the extra vector bosons present in these models
were not heavy enough (we deduce their masses in 
Section 4), they would introduce unacceptable
mixing with the EW gauge bosons and four fermion 
operators upsetting LEP2 data and atomic parity 
experiments\footnote{These terms would be absent
in models with a $T$ parity \cite{Cheng:2003ju}
or a smaller gauge group \cite{Bai:2008cf}}. 
This forces that $f$ must be large, above 3 TeV 
\cite{Schmaltz:2004de,Casas:2005ev,Han:2005dz,Marandella:2005wd}. 
However, this can be achieved with just one large 
VEV, $f_2\ge 3$ TeV, leaving 
$f_1$ unconstrained. We will then assume that
$f_2$ is large and consider values of $f_1$ between
$v/\sqrt{2}=174$ GeV and 1 TeV. 

The global symmetry is not exact (see below), so
the Higgs boson will get a one-loop potential and a VEV,
\beq
\langle h^0 \rangle = u/\sqrt{2}\;.
\eeq 
Such VEV implies the triplet VEVs
\beq
\langle \phi_{1}\rangle = 
\left(\begin{array}{c} i f_1 s_1 \\ 0 \\ f_1 c_1  \end{array}\right)
\; , \;\;\;\;
\langle \phi_{2}\rangle = 
\left(\begin{array}{c} -i f_2 s_2 \\ 0 \\ f_2 c_2 \end{array}\right)\;,
\eeq
where
\beq
s_1\equiv \sin {u f_2\over \sqrt{2} f f_1}
\; , \;\;\;\;
s_2\equiv \sin {u f_1\over \sqrt{2} f f_2}
\;.
\eeq
Since the two upper components in the triplets transform as
an $SU(2)_L$ doublet, it is clear that 
to obtain the observed $W$ and $Z$ masses one needs
\beq
\sqrt{f_1^2 s_1^2 + f_2^2 s_2^2} = {v\over \sqrt{2}} \approx 
174\;{\rm GeV}\;.
\eeq
In the limit with $f\approx f_2\gg f_1$ that we are considering
$s_1$ may be large, as $s_1\approx v/( \sqrt{2} f_1)$.

An important observation here is the following. 
Once $h^0$ gets the VEV $u/\sqrt{2}$ and  
we expand it (in the unitary gauge) as $h^0=(u+h)/\sqrt{2}$,  
we obtain that the physical Higgs $h$ has both
doublet and singlet components (in the first and third entries
of the triplet, respectively):
\beq
\phi_{1} \approx 
\left(\begin{array}{c} if_1 (s_1 
\cos \displaystyle{h \over \sqrt{2}f_1}
+ c_1\sin \displaystyle{h\over\sqrt{2} f_1}) \\ 0 \\ 
f_1 (c_1 \cos \displaystyle{h \over \sqrt{2} f_1} - s_1
\sin \displaystyle{h\over\sqrt{2} f_1}) \end{array}\right)
= 
\displaystyle{1 \over \sqrt{2}}
\left(\begin{array}{c} i c_1 h \\ 
0 \\ - s_1  h \end{array}\right)
+ ...
\label{higgs}
\eeq
If $f_1$ is much larger than the EW scale, $s_1$ is small
and $h$ is predominantly a doublet. However, as $f_1$ 
aproaches $v/\sqrt{2}$ the singlet component $s_1$ grows.
Where did the doublet component go? It is easy to see that
it went to the scalar $r_1$
that gets massive at $f_1$. 

This is a generic feature in any LH model. The scale $f$ of
global symmetry breaking is always defined by the VEV of an 
$SU(2)_L$ singlet, that gets a mass of order $f$. Then the EW 
symmetry breaking mixes the Higgs $h$ (the pseudo-GB of the 
global symmetry) with this massive singlet. 
Since the singlet component of $h$ does not couple,  
both its gauge couplings $g$ and $g'$ and its Yukawa 
couplings $\sqrt{2} m_f/v$ will appear suppressed by a 
factor of $c_1$.
These anomalous LH couplings have nothing to do with the 
non-linear realization
of the pseudo-GBs, they just reflect the mixing with the 
scalar singlet
massive at the scale of global symmetry breaking.
In our case, since $f_1$ can be close to $v/\sqrt{2}$
while consistent with all precision data, 
the effect may be large and observable 
at the LHC \cite{Barcelo:2007if}.

Notice also that the Little Higgs $h$, not being a pure
doublet, only unitarizes {\it partially} the SM cross sections 
involving massive vector bosons. In particular, the 
cutoff at $\approx 1.7$ TeV set by $WW$ elastic scattering would be
{\it moved} up to (1.7/$s_1$) TeV. Below that scale the massive
scalar $r_1$ (or other field) should complete the unitarization.

A final comment concerns the limit $f_1\rightarrow v/\sqrt{2}$.
The pseudo-GB $h$ becomes there a pure $SU(2)_L$ singlet, 
and the (unprotected) field $r_1$, massive at the scale $f_1$,  
becomes a doublet and is the {\it real} Higgs that breaks the 
EW symmetry. 
In this limit the naturality cutoff would be the same as in  
the SM, whereas in the general case
with $f_1>v/\sqrt{2}$ it is at $\approx 4\pi f_1$.

\section{Fermion masses}

Let us start discussing the top quark Yukawa sector. 
Since at the scale $f$ the local symmetry is $SU(3)\times U(1)_X$, 
we must include the doublet $Q^T=(t\; b)$ in a
triplet $\Psi_Q^T=(t\; b\; T)$, together with  
two singlets, $t^c_1$ and $t^c_2$. The Lagrangian may then contain
the four couplings 
\beqa
-{\cal L}_t &=& \lambda_1\; \phi_1^\dagger \Psi_Q t_1^c + \lambda_2
\;\phi_2^\dagger \Psi_Q t_2^c + \nonumber \\
&& \lambda'_1\; \phi_1^\dagger \Psi_Q t_2^c + \lambda'_2
\;\phi_2^\dagger \Psi_Q t_1^c + {\rm h.c.}\;,\label{yt}
\eeqa
where all the fermion fields are two-component spinors.
We will assume that only $\lambda_1$ is of order one and that
the rest of the couplings 
are one order of magnitude smaller. This could be justified 
if the global $SU(3)_1\times SU(3)_2$ symmetry is 
{\it approximate} in this sector.
If $\Psi_Q$ is a triplet under $SU(3)_1$, then the 
terms $\lambda_1$ and $\lambda'_1$ will be unsuppressed
(symmetric),
whereas  $\lambda_2$ and $\lambda'_2$ break the symmetry and
will be smaller. We can then redefine
the fields $t^c_{1,2}$ so that $\lambda_1\rightarrow
\sqrt{\lambda_1^2+{\lambda'_1}^2}$ and $\lambda'_1\rightarrow 0$,
{\it i.e.}, with all generality we can take $\lambda'_1=0$
and $\lambda_2$ and $\lambda'_2$ small.  

In the original simplest LH model
$\lambda'_{1,2}=0$ and $\lambda_{1,2}$ 
break collectively the symmetry ({\it i.e.,}
only in diagrams that contain simultaneously both couplings)
\footnote{If $\lambda'_{2}=0$ two-loop diagrams generate
corrections of order $\lambda_1^3\lambda_2^4/(16\pi^2)^2$.}. 
Here the diagrams that only involve the large 
top-quark Yukawa coupling ($\approx \lambda_1$)  
do not break the symmetry, and this is enough 
to rise the natural cutoff of the SM above LHC energies.

Keeping the exact dependence on $f_{1,2}$, on the Higgs
VEV $u$, and on the possible $CP$-odd singlet VEV 
$\langle \eta \rangle=y$, and performing appropriate phase
redefinitions of the fermion fields, we obtain the mass matrix 
\beq
-{\cal L}_t \supset  
\left( \begin{array}{cc} t & T \end{array} \right)
\left(\begin{array}{cc} 
\lambda_1 f_1 s_1-e^{i\theta} \lambda'_2 f_2 s_2&
-\lambda_2 f_2 s_2+e^{-i\theta} \lambda'_1 f_1 s_1\\
\lambda_1 f_1 c_1+e^{i\theta} \lambda'_2 f_2 c_2&
\lambda_2 f_2 c_2+e^{-i\theta} \lambda'_1 f_1 c_1
\end{array} \right)
\left(\begin{array}{c} t_1^c \\
t_2^c
\end{array} \right)\;,
\eeq
where 
\beq
\theta= {y f\over \sqrt{2}f_1 f_2}
\eeq 
and $\lambda_{1,2}$ are both real.
Several comments are here in order. 
First, the mass of the 
extra $T$ quark is just
\beq
m_T^2=(\lambda_1^2+{\lambda'_1}^2)\; f_1^2 + 
(\lambda_2^2+{\lambda'_2}^2)\; f_2^2 
+2(\lambda_1\lambda'_2+\lambda_2{\lambda'}_1)\;f_1f_2
\;c_{12} c_\theta -m_t^2\;,
\eeq
with 
\beq
c_{12} \equiv \cos {u f\over \sqrt{2} f_1 f_2}\;.
\eeq
Since all the couplings except for $\lambda_1$ are
small, and this coupling only contributes to $m_T$
multiplied by the lower VEV $f_1$, the extra $T$ quark
will have a mass of order $f_1$. 
Second, notice that if $\lambda'_{1,2}=0$ (the 
collective breaking case) then the fermion masses do not
depend on the value $y$ of the singlet $\eta$. As
a consequence, $\eta$ will not get an effective potential
and will remain massless at that order. We show in
the next section that in this case the Higgs mass is
always below present bounds from LEP \cite{Barate:2003sz}.

The smaller up and charm quark masses could appear if the
assignments for the quark triplets under the 
approximate symmetry are different: 
triplets under the second $SU(3)_2$, 
singlets under the first one. In particular,
the only large Yukawas (one per family) should couple 
these triplets with $\phi_2$. That would make the 
extra up-type
quarks very heavy ($m_{C,U}\approx f_2$), whereas the 
up and the charm fields would couple to the Higgs with
suppressed Yukawa couplings.

Down-type quarks (and also charged leptons) 
may get their mass through 
dimension 5 operators \cite{Schmaltz:2004de} like
\beq
-{\cal L}_b \approx
{y_b\over f}\; \phi_1 \phi_2 \Psi_Q b^c + {\rm h.c.}\;,
\eeq
but they do not require extra fields not large couplings.

Finally, here
the lepton doublets become triplets that include
a $SU(2)_L$ singlet: $\psi_L^T=(\nu\;e\;N)$. This forces 
the addition of a fermion singlet $n_c$ per family and the 
Yukawa couplings
\beq
-{\cal L}_\nu = {\lambda'_1}^\nu\; \phi_1^\dagger \Psi_L n^c +
\lambda_2^\nu\; \phi_2^\dagger \Psi_L n^c 
+ {\rm h.c.}\;.\label{ynu}
\eeq
The approximate symmetry should imply then
$\lambda^\nu_2\approx 1 \gg {\lambda'_1}^\nu$, 
and the two extra fermions ($N$ and $n^c$) 
would combine into a Dirac field
of mass $\approx f_2$. For the light neutrinos, 
in these models there is an alternative to
the usual see-saw mechanism. In 
\cite{delAguila:2005yi} it is shown that a small lepton
number violating mass term
\beq
-{\cal L}_\nu \supset {1\over 2} m\; n^c n^c
+ {\rm h.c.}
\eeq
of order 0.1 keV would generate a one-loop neutrino 
mass $m_\nu\approx 0.1$ eV.

In summary, in this model all the extra right-handed neutrinos
and up-type quarks except for the one cancelling 
top-quark quadratic corrections can be very heavy, with 
masses around $f_2\approx 3$ TeV. 
As for the extra $T$ quark, 
all precission bounds are respected if its
mixing $V_{Tb}$ with the standard top quark is smaller
than $\approx 0.2$ \cite{Aguilar-Saavedra:2002kr,Barcelo:2007if}.

\section{Gauge boson masses}

The $SU(3)\times U(1)_X$ 
gauge boson masses come from terms 
$(D^\mu\Phi_i)^\dagger (D_\mu\Phi_i)$ in the Lagrangian. In the 
charged sector we have
\beq
D_\mu\phi_1 \supset -ig \sum_{i=1,2,6,7} A^i_\mu T^i \phi_1
={g f_1\over \sqrt{2}}
\left( \begin{array}{c} 0 \\ 
s_1W_\mu-c_1W'_\mu \\ 0
\end{array} \right)\;,
\eeq
where we have defined
\beq
W={1\over \sqrt{2}}\left( A^1_\mu+i A^2_\mu \right)\;;\;\;
W'={1\over \sqrt{2}}\left( A^7_\mu+i A^6_\mu \right)\;,
\eeq
with an analogous expression for $D_\mu\phi_2$. 
In this basis the mass matrix reads  
\beq
{g^2\over 2}
\left( \begin{array}{cc} f_1^2 s_1^2+f_2^2 s_2^2&
f_2^2 s_2 c_2 - f_1^2 s_1 c_1\\ 
f_2^2 s_2 c_2 - f_1^2 s_1 c_1 & f_1^2 c_1^2+f_2^2 c_2^2
\end{array} \right)\;, 
\eeq
and has the eigenvalues 
\beq
M^2_{W_1(W_2)}={g^2f^2\over 4} \left(
1 -(+) \sqrt{1- s_{2\beta}^2\; s_{12}^2}\right)\;,
\eeq
where
\beq
s_{2\beta}\equiv 2 {f_1 f_2 \over f^2}\;.
\eeq
Notice that the two masses add to a constant
independent of the Higgs VEV (in $s_{1,2}$), 
which will imply no quadratic
divergencies in the potential at the one-loop level.

In the neutral sector we find 
\beqa
D_\mu\phi_1 &\supset & \left(-ig \sum_{i=3,4,5,8} A^i_\mu T^i \phi_1
+ {i g_X \over 3} A^X_\mu \right)\nonumber \\
&=&{g f_1\over 2}
\displaystyle 
\left( \begin{array}{c} s_1 \sqrt{1+t^2} Z_\mu + s_1 
\displaystyle{1-t^2\over
\sqrt{3-t^2}} Z'_\mu - c_1 A^5_\mu -i c_1 A^4_\mu\\ 
0 \\ 
s_1 A^4_\mu + i s_1 A^5_\mu + i 2 c_1 
\displaystyle {1\over \sqrt{3-t^2}} Z'_\mu 
\end{array} \right)\;,
\eeqa
where
\beq
t={g'\over g}= \sqrt{3}\; {g_X\over \sqrt{3g^2+g_X^2}}
\eeq
and
\beqa
Z'_\mu&=&\sqrt{1-{t^2\over 3}}\; A^8_\mu +
{t\over \sqrt{3}}\; A^X_\mu\;,\nonumber  \\
Z_\mu&=&{1\over \sqrt{1+t^2}}\left( A^3_\mu 
+{t^2\over \sqrt{3}}\; A^8_\mu
-t \sqrt{1-{t^2\over 3}}\; A^X_\mu \right)\;.
\eeqa
The mass matrix of $(Z_\mu\;,Z'_\mu\;,A^4_\mu\;,A^5_\mu)$
is then 

\beq
{g^2\over 2}
\left( \begin{array}{cccc} (1+t^2) (f_1^2 s_1^2+f_2^2 s_2^2)&
{(1-t^2)\sqrt{1+t^2}\over \sqrt{3-t^2}}(f_1^2 s_1^2+f_2^2 s_2^2)&
0&-\sqrt{1+t^2}(f_1^2 s_1c_1-f_2^2 s_2 c_2)\\
{(1+t^2)\sqrt{1+t^2}\over \sqrt{3-t^2}}(f_1^2 s_1^2+f_2^2 s_2^2)&
\begin{array}{c} \\
{(1-t^2)^2\over 3-t^2}(f_1^2 s_1^2+f_2^2 s_2^2)+\\
{4\over 3-t^2}(f_1^2 c_1^2+f_2^2 c_2^2)\\ 
\end{array}&
0&{1+t^2\over \sqrt{3-t^2}}(f_1^2 s_1c_1-f_2^2 s_2 c_2)\\
\begin{array}{c} 0 \\ \; \end{array}&0&f^2&0\\
-\sqrt{1+t^2} (f_1^2 s_1c_1-f_2^2 s_2 c_2)&
{1-t^2\over \sqrt{3-t^2}}(f_1^2 s_1c_1-f_2^2 s_2 c_2)&
0& f^2
\end{array} \right)\;.
\eeq
It is easy to see that, again, the trace of this matrix does not
depend on the Higgs VEV:
\beq
{\rm Tr}\;[ M^2 ] = {g^2\over 2} \left( {4\over 3-t^3}+2 \right) f^2\;.
\eeq
The mixing terms of $A^5_\mu$ with $Z_\mu$ and $Z'_\mu$ were
overlooked in \cite{Schmaltz:2004de}. Although they cancell
at the lowest order 
in $v/(\sqrt{2} f)$, we show in the next
section that they are essential to obtain the right 
ultraviolet (UV) dependence of the effective potential. 

\section{One-loop potential from collective breaking}

Gauge and Yukawa couplings
break in this model the global symmetries and
introduce a one-loop potential for the pseudo-GBs. 
To be realistic,
we need that the potential implies the right Higgs VEV and 
an acceptable mass for $h$.
The potential 
can be given in terms of the fermion and boson masses
expressed as a function of the Higgs. We will consider
here the contributions from the top-quark (the rest of
Yukawas do not introduce any new effects) 
and the gauge sectors. 

From the
top-quark sector we have
\beq
V_{top}=-{3\over 16\pi^2} \Lambda^2\; {\rm Tr}\; [m^\dagger m]+
{3\over 16\pi^2}\; {\rm Tr}\; [(m^\dagger m)^2 \log \left( 
\Lambda^2\over m^2 \right) ]
\label{vtop}
\eeq
Let us first discuss the collective breaking case, with
$\lambda'_{1,2}=0$ in Eq.~(\ref{yt}). The two mass eigenvalues 
are just 
\beq
m_{t(T)}^2(h)={M^2\over 2} \left(
1 -(+) \sqrt{1-s_{2\alpha}^2\; s_{12}^2(h)}\right)\;,
\eeq
with 
\beqa
s_{12}(h)& = & \sin {h f\over \sqrt{2} f_1 f_2}
\;;\nonumber \\
M^2 & = & \lambda_1^2 f_1^2 + \lambda_2^2 f_2^2
\;;\nonumber \\
s_{2\alpha}& = & {2\lambda_1 \lambda_2 f_1 f_2\over 
\lambda_1^2 f_1^2 +\lambda_2^2 f_2^2}
\;.
\eeqa
$V_{top}$ presents in this case several important features. First,
$m_t^2+m_T^2$ is a constant (does not depend on $h$),
so the quadratic divergence (first term in (\ref{vtop})) 
is zero. Second, 
we can write (up to a constant) 
\beq
V_{top}=
{3\over 16\pi^2} m_t^4 \log \left( 
m_T^2\over m_t^2 \right) +
{3\over 16\pi^2} \left( m_t^4+m_T^4\right) \log \left( 
\Lambda^2\over m_T^2 \right)\;.
\eeq
As noticed in \cite{Schmaltz:2005ky}, this potential
can be understood as the usual quartic up-quark 
correction below $m_T$, plus an 
$SU(3)$-symmetric correction proportional to
\beq
m_t^4+m_T^4={M^4\over 2} \left( 2 
- s^2_{2\alpha} \right) \; s_{12}^2(h) 
\eeq
above that scale. This second contribution is 
logarithmically divergent, and it would  
redefine (renormalize) the quartic
\beq
V_{UV}=a\; \left( \phi_1^\dagger
\phi_2 \right)_{\bf 1}
\left( \phi_2^\dagger \phi_1 \right)_{\bf 1}\supset
a\;  f_1^2 f_2^2 \;\left( 1- s_{12}^2(h)\right)\;.
\label{Vuv}
\eeq
The {\it sensitivity} of the potential to the physics 
in the UV can be accounted by taking $a$ as a free 
parameter or, equivalently, setting $a=0$ and varying freely 
the cutoff. We will take this second approach, defining 
in this sector an arbitrary cutoff $\Lambda_t$ 
that may be different from the one
in the gauge sector, $\Lambda_g$. Notice also that any
UV contribution to the (adimensional) parameter $a$ 
should be {\it small}, as this parameter 
breaks the global symmetry. 

Let us analize now the gauge sector. The one-loop contribution
to the effective potential is
\beq
V_{gauge}={3\over 64\pi^2} \Lambda^2_g {\rm Tr}\; [M^2]+
{3\over 64\pi^2} {\rm Tr}\; [ M^4 \log \left( 
\Lambda^2_g\over M^2 \right) ]
\eeq
Again, {\it (i)} the quadratic divergence vanishes, 
{\it (ii)} below the scale
$\approx g f$ of the massive vector bosons one has the usual
$W^\pm,Z$  corrections, and {\it (iii)} above that scale 
there is an $SU(3)$-symmetric 
logarithmic divergence proportional to the sum of all 
vector bosons masses to the fourth power. 
In particular, in the charged sector  
($W_i$ carries particle plus antiparticle)
\beq
M^4_{W_1}+M^4_{W_2}={g^4 f^4\over 4} \left(
1 - {1\over 2} s_{2\beta}^2\; s_{12}^2(h)\right)\;, 
\eeq
whereas the four neutral vectors give 
\beq
\sum_{i=1}^4 M_{Z_i}^4  = {g^4 f^4\over 2} 
\left( 1+{8\over (3-t^2)^2} - {1+t^2\over 3-t^2} s_{2\beta}^2\; 
s_{12}^2(h) \right)\;.
\eeq
These divergent terms will renormalize a combination of
the operator in (\ref{Vuv}) and 
\beq
V'_{UV}=b\; \left( \phi_1^\dagger
\phi_2 \right)_{\bf 8}
\left( \phi_2^\dagger \phi_1 \right)_{\bf 8}\supset
{2b\over 3}\;  f_1^2 f_2^2 \;\left( 2+ s_{12}^2(h)\right)\;.
\label{q88}
\eeq
We find that the UV physics (quartic terms proportional to
$a$ and $b$ {\it or} the cutoffs $\Lambda_{t,g}$ in the 
top-quark and gauge sectors) can only define 
the coefficient of a term proportinal to $s_{12}^2(h)$. 
This single arbitrary parameter from the UV completion will 
not be enough (see below) to obtain an acceptable Higgs mass.

Let us fix $f_2$ at 3 TeV and vary $f_1$ between 200 GeV 
and 1 TeV. In the model with collective breaking ({\it i.e.}, 
$\lambda'_{1,2}=0$ in the top-quark sector) the effective Higgs
potential will change with the values of $\lambda_{1,2}$ and
the cutoff $\Lambda_t$ (as explained before, 
the potential is only sensitive to a combination of 
$\Lambda_t$ and $\Lambda_g$, so we fix $\Lambda_g$ at 5 TeV).
These three parameters must produce $M_Z=91$ GeV  
({\it i.e.}, the right Higgs VEV) and $m_t=171$ GeV.
We will require that 
the extra $T$ quark has a mass below 2 TeV 
(in order to cancell {\it naturally} top-quark quadratic 
corrections), and 
that its mixing $V_{Tb}$ with the top is smaller than 
0.25. In Fig.~1 we plot the maximum value of the Higgs mass 
for different values of $f_1$ and any consistent value of the
other parameters. All these values of $m_h$ are far below 
the LEP bound of 121 GeV \cite{Barate:2003sz}, a fact that 
does not change increasing $f_2$.
\begin{figure}
\begin{center}
\includegraphics[width=90mm]{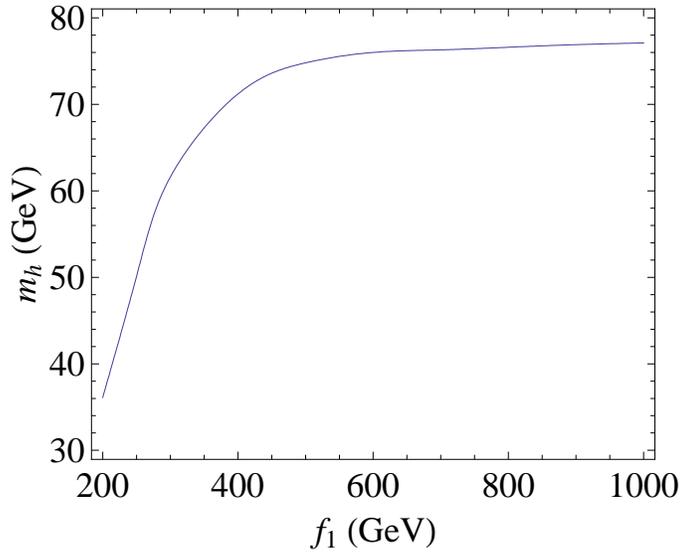}
\caption{Higgs mass in the collective breaking
case for $f_2=3$ TeV and different values of $f_1$.
\label{fig1}}
\end{center}
\end{figure}

\section{Effective potential in the minimal model}

As described in Section 2, we propose a model with
$f_1\approx 0.1 f_2$, $\lambda_1\approx 1$, and 
the rest of the couplings in the top quark sector at least
one order of magnitude smaller.
The model allows heavy extra gauge bosons while the 
$T$ quark that cancels top corrections can be below
1 TeV. Notice that, since the global symmetry in 
the top sector is
approximate, the cancelation of one-loop quadratic 
corrections will also be approximate. This 
suffices to increase the {\it natural} cutoff 
of the model from 1 TeV 
up to 5--10 TeV. On the other hand, in the one-loop
effective potential 
there will be new operators breaking the 
global symmetry that do not appear in the 
collective breaking case.
As a consequence, the Higgs will get a mass 
above LEP bounds, and also the scalar singlet 
$\eta$ will acquire an acceptable mass.

The suppressed couplings $\lambda'_{1,2}$ in the 
top-quark Yukawa sector imply that $\Phi_Q t_1^c$ 
couple both to $\Phi_1$ and $\Phi_2$, introducing 
the one-loop quadratic divergence 
\beqa
\Delta V_{top}=-{3\over 16\pi^2} \Lambda_{t}^2  
\biggl( &  \!\!\!\!\!\!\!\!\!\!\!\!\!\!\!\!
\!\!\!\!\!\!\!\!\!\!\!\!\!\!\!\!\!\!\!\!\!\!\!\!
f_1^2 \left( \lambda_1^2 + {\lambda'}_1^2 \right) + 
f_2^2 \left( \lambda_2^2 + {\lambda'}_2^2 \right) + \cr
& 2 f_1 f_2 
\left( \lambda_1 \lambda'_2 +\lambda_2 \lambda'_1\right)
\cos \displaystyle{h f\over \sqrt{2} f_1 f_2}\; 
\cos \displaystyle{\eta f\over \sqrt{2}f_1 f_2}\; \biggr) \;.
\eeqa
This term is determinant in order to obtain an 
acceptable potential because it is proportional 
to $c_{12}(h)$, while before all the UV-dependent contributions
(both from the top-quark and the gauge sectors) were proportional
to $s_{12}^2(h)$ ({\it i.e.}, $c_{12}^2(h)$). 
To illustrate that, we consider a particular
set of parameters for $f_1=400$ GeV and $f_2=3$ TeV. 
We take $\lambda_1=1.19$, $\lambda_2=-0.25$, $\lambda'_2=0.03$ 
and $\lambda'_1=0$ (see Section 3). 
We fix the UV cutoff in the top-quark sector to 
$\Lambda_t=\Lambda_{g}=5$ TeV, but we include 
the UV dependent coupling $a$ in Eq.~(\ref{Vuv}) with the
value $a=1/(16\pi^2)$.
\begin{figure}
\begin{center}
\includegraphics[width=100mm]{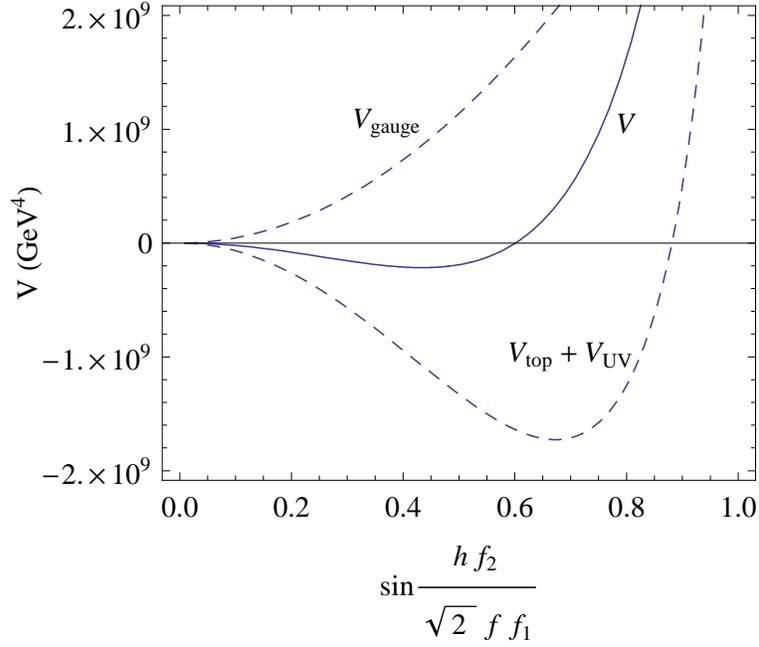}
\caption{One-loop Higgs potential as a function of $s_1(h)$ 
for the choice of parameters given
in the text.
\label{fig2}}
\end{center}
\end{figure}
In Fig.~2 we separate the contributions from  
the top and the gauge sectors (we have added $V_{UV}$ to 
$V_{top}$). We
plot the potential expressed as a function of
$s_{1}(h)$. In the minimum, $s_{1}=0.43$ (i.e., $u=259$ GeV),
which reproduces the values $M_Z=91$ GeV and $m_t=171$ GeV, 
with an extra $T$ quark of 920 GeV. This potential implies
$m_h=156$ GeV and $m_\eta=107$ GeV. Increasing
the value of $f_1$, changing the 
value of $\lambda'_2$, and varying
the parameter $a$ we obtain Higgs masses above 200 GeV.

In order to see if the solutions that we find involve any
amount of fine tuning, we have varied in a $\pm 5\%$ the VEV 
$f_1$, the large Yukawa coupling $\lambda_1$, and the 
UV-dependent coupling $a$ from the values given above. 
\begin{figure}
\begin{center}
\includegraphics[width=100mm]{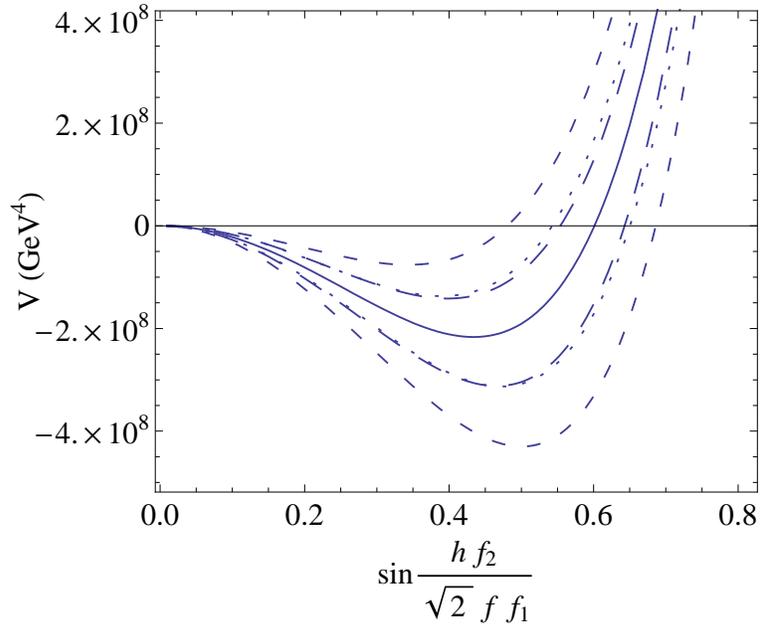}
\caption{Variation of the Higgs potential 
for a $\pm 5\%$ variation of $f_1$ (short dashes)
$\lambda_1$ (long dashes) and $a$ (dots)
versus the central values given in the text.
The EW scale $v/\sqrt{2}$ changes in up to a 
$+20\%$ or a $-25\%$, whereas $m_h$ varies 
between 126 and 178 GeV.
\label{fig3}}
\end{center}
\end{figure}
In Fig.~3 we plot the changes in the Higgs potential 
caused by the variation of 
each one of these parameters. We obtain 
that the EW scale $v/\sqrt{2}$ changes between $+20\%$
and $-25\%$ respect to the central values, whereas the 
Higgs mass moves between 126 and 178 GeV. This result shows 
that the Higgs sector of the model does not involve any 
severe degree of fine tuning (just $1-5/20 =75\%$ 
cancellations). 

\section{Summary and discussion}
The Higgs doublet may be the pseudo-GB of a global symmetry 
broken spontaneously at a higher scale $f$. 
The EW symmetry breaking will then define 
a physical Higgs $h$ that has $SU(2)_L$ doublet 
and singlet components, 
while a scalar singlet of mass $\approx f$ acquires 
a doublet component. LH models include  
an extra $T$ quark that cancells top-quark 
quadratic corrections to the Higgs bilinear. 
This cancellation is effective
(making the model natural) if the mass of the
$T$ quark is below 1 TeV. On the other hand, the
extra gauge bosons present in these models also get 
a mass of order $f$, which may conflict precision 
EW data if $f<3$ TeV. 

We have shown that Schmaltz's {\it simplest} LH model can 
naturally accommodate heavy gauge bosons, a lighter $T$
quark, and an acceptable Higgs mass 
if {\it (i)} $f_1\approx 0.1 f_2$ and  
{\it (ii)} in the top-quark sector there is only one
large Yukawa coupling, 
$\lambda_2, \lambda'_{1,2}\lsim 0.1 \lambda_1$. The 
second condition makes the global symmetry approximate,
in contrast with the usual scenario with 
$\lambda_1\approx \lambda_2$ and 
$\lambda'_{1,2}=0$.

We have studied in detail the Coleman-Weinberg potential
and have shown that under these two conditions the model
gives {\it naturally} acceptable EW minima with a Higgs 
mass above present bounds. 
It is essential to work at all order
in $v^2/f^2$ 
(we express the results in terms of the sine of this ratio), 
as the usual first order expansion fails in
the cases with low $f_1$ considered here. In particular,
we have found that in the collective breaking case with 
$\lambda_{1,2}=0$ it is
impossible\footnote{The authors in \cite{Cheung:2006nk}
find acceptable cases working at first order in $v^2/f^2$} 
to obtain a Higgs mass above LEP bounds.
The basic reason is that both top-quark and gauge corrections
give a logarithmic divergent term proportional to 
$s_{12}^2(h)$. This fact is not apparent in the 
calculation of the potential at first order in $h^2/f^2$
given in \cite{Schmaltz:2004de}, 
as the author overlooks the mixings of $A^5_\mu$ 
($W'_{0,\overline 0}$ there) with 
the other neutral gauge bosons. We have shown that
in the framework with $f_1\ll f_2$ it is natural to
have an approximate global symmetry in the Yukawa 
sector, since just one large coupling per flavour 
is enough to generate the top-quark mass and masses above
the TeV for all the extra fermions but $T$. This also 
provides an acceptable Higgs mass with no need for
a $\mu$ term $\phi_1^\dagger \phi_2$ put by hand in 
the scalar potential. In addition, we showed 
that the mass of the pseudoescalar $\eta$ 
may be here larger than in the usual scenario with collective
breaking and an extra $\mu$ term, closing 
(or reducing) kinematically the 
Higgs decay channel $h\rightarrow \eta\eta$. 
This decay mode could provide very interesting signals
in a hadron collider when the Higgs is produced together with a
$W$ or a $Z$ gauge boson \cite{Cheung:2007sva}.

The framework discussed here is a minimal departure from the 
SM with features that should be observed 
at the LHC for the preferred 
values of $f_1$ below 500 GeV. 
The Higgs appears mixed with a singlet that may be as
heavy as $4\pi f_1$ (when the LH is completed with
strongly interacting physics \cite{Giudice:2007fh}). 
The observation of its 
anomalous gauge or fermion couplings, and/or the 
observation of a vectorlike $T$ quark, would then 
reveal the new scale $f_1$. The resulting model, with
a natural cutoff higher than the one in the SM but
right above LHC energies, would
certainly be an invitation to plan for a bigger
collider.

\section*{Acknowledgments}
This work has been partially supported by
MEC (FPA2006-05294) and by Junta de Andaluc\'{\i}a
(FQM 101 and FQM 437).

\end{document}